\documentstyle{article}

\title{Entropy production by resonance decays\\
\vspace{0.5cm}
{\normalsize TPR 96--09,\ DUKE--TH--96--117}}

\author{    
  Stefan Ochs\\
   Institut f\"ur Theoretische Physik, Universit\"at 
   Regensburg, D-93040 Regensburg, Germany\\
   \vspace{1.cm}\\
   Ulrich Heinz\thanks{On sabbatical leave from Institut f\"ur Theoretische
     Physik, Universit\"at Regensburg, D-93040 Regensburg, Germany.}\\ 
   Physics Department, Duke University, Durham, NC 27708-0305, USA
}

\renewcommand{\textfraction}{0.0}

\input epsf.tex           % reactivate to print postscript figures

\begin{document}
  \hyphenation{}
  \maketitle

  \begin{center}
  \end{center}

 \begin{abstract}
    We investigate entropy production for an expanding system of 
    particles and resonances with isospin symmetry -- in our case 
    pions and $\rho$ mesons -- within the framework of relativistic 
    kinetic theory. A cascade code to simulate the kinetic equations 
    is developed and results for entropy production and particle 
    spectra are presented.  
 \end{abstract}

\section{Introduction}

The search for the Quark-Gluon-Plasma (QGP) requires a detailed and 
quantitative understanding of the influence of known physics on the 
particle spectra and abundances from heavy ion collisions, such that 
possible systematic deviations due to the new physics of the QGP can 
be recognized in the experimental data.  

When interpreting the pion momentum spectra 
\cite{Sollfrank92a}-\cite{Schned94}
the question of entropy production during freeze-out occurred 
\cite{Sollfrank92b}. If the contribution to entropy production 
after freeze-out due to resonance decays is known, the entropy of the 
system before freeze-out can be calculated by measuring the 
multiplicities and momentum spectra of all particles after freeze-out 
and combining these data with an estimate of their initial spatial 
distribution (e.g. from a fireball model). It is then crucial to know 
how much entropy is produced in an expanding system during the 
transition (freeze-out) from a highly interacting (equilibrium) state 
to a non-interacting one. Here we will give such an estimate for a 
simple system of pions and $\rho$ mesons, using the framework of 
relativistic kinetic theory \cite{degroot}. This will enable us to 
estimate the reliability of models based on hydrodynamics in which 
freeze-out is usually implemented by cutting off the 
entropy-conserving expansion sharply at the so-called freeze-out 
surface, neglecting possible entropy production during the decoupling 
process or afterwards by the decay of unstable resonances.  

The paper is organized as follows:
In Section \ref{kin} we will formulate via relativistic kinetic theory 
the problem of resonance formation and decay for a system of pions and 
$\rho$ mesons, including Bose statistical effects. We will show that
the collision term satisfies Boltzmann's H-theorem and discuss the 
formation and decay 
rates for $\rho$ mesons. In Section \ref{sim} we describe a simulation 
of the kinetic equations of our system with the help of a cascade 
code, and in Section \ref{res} we present our results, in particular 
on the production of entropy by resonance formation and decays. 
We will also discuss the influence of the interplay between 
$\rho$-formation and -decay on the time evolution of the particle
spectra.
In Section \ref{con} we summarize our results. Some technical details
are given in the Appendix.  

In the following we will use natural units, $\hbar = c = k_{\rm B} = 
1$, if not stated explicitly otherwise.

\section{Kinetic theory for the $\pi$-$\rho$-system}
\label{kin}

As a starting point we consider the kinetic equation
 \begin{equation} 
 \label{kinequ}
   m \frac{d}{d\tau} f(x,p) = - \Gamma_{\rm loss}(x,p) \ f(x,p) + 
   \Gamma_{\rm gain}(x,p) \  \bigl( 1 + f(x,p) \bigr),  
 \end{equation}
where $\tau$ is the proper-time of a particle at point $x$ with 
momentum $p$, and the right hand side  of Eq.~(\ref{kinequ}) is 
the collision term. The latter consists of a term for particle loss 
(with  rate $\Gamma_{\rm loss}$) and one for particle gain (with rate 
$\Gamma_{\rm gain}$). A vanishing collision term means vanishing 
entropy production (Liouville theorem). The left hand side of the 
kinetic equation can be rewritten as  
 \begin{equation} 
 \label{grgl2}
   m \frac{d}{d\tau} = 
   \left( m \frac{dx^{\mu}}{d\tau} \right) \frac{\partial}{\partial x^{\mu}} 
   + \left( m \frac{d p^{\mu}}{d \tau} \right) 
   \frac{\partial}{\partial p^{\mu}}
   =  p^{\mu} \frac{\partial}{\partial x^{\mu}} + F^{\mu} 
   \frac{\partial}{\partial p^{\mu}}.
 \end{equation}
In the following we neglect external or mean-field forces by setting 
$F^{\mu}$ to zero.  

For $2 \rightarrow 2$ processes Boltzmann's H-theorem can be proven 
from unitarity and particle number conservation alone \cite{degroot} 
whereas in $2 \rightleftharpoons 1$ reactions as in resonance 
formation and decay a different approach has to be taken. In 
\cite{Stanis86} the bilateral normalization condition was exploited.  
We will here consider the isospin symmetry of the matrix elements to 
evaluate the gain and loss rates for the $\pi$-$\rho$-system and thus 
prove the Boltzmann H-theorem explicitly.  

At low relative momentum the $\pi\pi$ interaction is dominated by the 
$\rho$ resonance, with a small non--resonant background \cite{weise}. We 
are interested in the very last stage of a nuclear collision, after 
the pions and $\rho$ mesons have ``frozen out", i.e. stopped interacting 
with other particles and with each other. After this freeze-out we 
expect the dominant physical process to be the decay of the unstable 
$\rho$ mesons which we want to describe by kinetic theory. But in 
order for such a description to be consistent and the calculation of 
entropy production to be meaningful, our collision term should allow 
for detailed balance, which implies that we also have to include the 
inverse process $\pi\pi\to\rho$ (resonant $\pi\pi$ scattering), 
however small its rate should be.

The interaction Hamiltonian for the $\pi$-$\rho$-system is of the form
\cite{weise}
 \begin{equation}
 \label{Hint}
   H_{\rm int} = f_{\rho \pi \pi}\ {\vec\rho}^{\,\mu} 
   \left(\vec\varphi \times
   \partial_{\mu} \vec\varphi \right),
 \end{equation}
where the vectors refer to the isospin structure, and $\vec{\rho}^{\, 
\mu}$ is a massive spin-1 vector field satisfying the gauge condition 
 \begin{equation}
   \partial_{\mu} {\vec\rho}^{\,\mu} = 0.
 \end{equation}
Due to energy momentum conservation and isospin symmetry only the 
processes $\rho^{\alpha} \leftrightarrow \pi^{\beta} \pi^{\gamma}$
are allowed where $\alpha,\beta,\gamma \in \{0,+,-\}$ are 
pairwise different. 

To write down the rates in terms of the distribution functions, we 
assume that the ensemble averages of multiple products of creation and 
annihilation operators arising from treating $H_{\rm int}$ in first 
order perturbation theory can be factorized into products of ensemble 
averages of 1-particle operators, given by the following 
replacement (for $\rho$ mesons $\alpha$ is a double index denoting 
both the spin and isospin component of the particle): 
 \begin{eqnarray}
   {\langle a^{\alpha \dagger} (k)\, a^{\alpha'} (k') \rangle} 
  &=&
   \delta(\vec k - {\vec k}')\, f_{\alpha} (k)\, \delta^{\alpha \alpha'}
  \\
   {\langle a^{\alpha} (k)\, a^{\alpha' \dagger} (k') \rangle}
  &=& 
   \delta(\vec k - {\vec k}')\, ( 1 + f_{\alpha} (k) ) \,
   \delta^{\alpha \alpha'}
  \\
   {\langle a^{\alpha} (k)\, a^{\alpha'} (k') \rangle}
  &=&
   {\langle a^{\alpha \dagger} (k)\, a^{\alpha' \dagger} (k')\rangle}
   = 0.
 \end{eqnarray}
In global thermal equilibrium this is justified by the thermal Wick 
theorem \cite{FW}, and in \cite{Sinyukov} it was 
shown how to 
generalize the latter to local thermal equilibrium systems. For 
non-equilibrium ensembles the validity of this ansatz has been proven 
from first principles only for the $\phi^4$-theory \cite{Stanis90}. We 
use it here because it leads to the classically expected result.  

Let us now introduce pion and $\rho$ meson distribution functions 
$f_\alpha \equiv f_{\pi_\alpha}$ and $F_\alpha \equiv 
f_{\rho_\alpha}$. In the case of the $\rho$, $F_\alpha$ describes a 
single isospin {\em and} spin projection, but we assume spin 
saturation such that the distribution functions for the three spin 
projections are equal, and therefore suppress the spin index 
on $F_\alpha$. The loss rates are then given by 
 \begin{equation}
 \label{pilossrate}
   \Gamma_{\rm loss}^{\pi^{\alpha}} (p_1) 
   = (2 \pi)^{4} {f_{\rho\pi\pi}^2}
    \sum_{\beta,\gamma}
    \vert \varepsilon_{\alpha \beta \gamma} \vert 
    \int Dp_2 Dp_3 
    \Bigl({\cal M} (p_1,p_3) + {\cal M} (p_{2},p_{3}) \Bigr)
    f_\beta (p_2) \left( 1 + F_{-\gamma} (p_3) \right) \,
    \delta (p_1 + p_2 - p_3) \, ,
 \end{equation}
 \begin{equation}
 \label{rholossrate}
    \Gamma_{\rm loss}^{\rho^{\alpha}} (p_3)
    = (2 \pi)^{4} 
    {f_{\rho\pi\pi}^2 \over 
    s_\rho(s_\rho+1)}
    \sum_{\beta,\gamma}
    \vert \varepsilon_{\alpha \beta \gamma} \vert 
    \int Dp_1 Dp_2
    \, {\cal M} (p_{1},p_{3})
    \left( 1 + f_{-\beta} (p_1) \right)
    \left( 1 + f_{-\gamma} (p_2) \right)
    \delta ( p_1 + p_2 - p_3), 
 \end{equation}
while the gain rates are simply given by replacing $f_\alpha 
\leftrightarrow 1 + f_\alpha$ and $F_\alpha \leftrightarrow 1 + 
F_\alpha$. Here $\varepsilon_{\alpha \beta \gamma}$ is the totally 
antisymmetric tensor, $Dp$ is the Lorentz-invariant measure in 
momentum space 
 \begin{equation}
   Dp := {d^4p \over (2\pi)^4}\, 2 \pi\, \delta(p^2 - m^2)
         \,\Theta(p_{0}) 
       = {1\over (2\pi)^3}\,{d^3p\over 2E} \, ,
 \end{equation}
where $E=\sqrt{{\vec p}^{\,2} + m^2}$, and the distribution functions 
are always taken on the mass-shell.

The two contributions to the pion loss term in Eq.~(\ref{pilossrate}) 
arise from the two possibilities in (\ref{Hint}) to let the derivative 
act on either one of the two scattering pions. 

The momentum dependent part of the matrix element as obtained from 
(\ref{Hint}) in first order perturbation theory is given by 
 \begin{equation}
 \label{matrixelement1}
   {\cal M} (p_i,p_\rho) =  - p_i^2 + \frac{( p_i p_\rho)^2}{m_\rho^2}
   = \frac{1}{m_\rho^2} 
     \left( \frac{1}{4} \left( (p_i \pm p_\rho)^2
            - p_i^2 - p_\rho^2 \right)^2
           - p_i^2 m_\rho^2 \right).
 \end{equation}
Here we have already summed over the three spin states of the $\rho$. 
To describe a single $\rho$ spin state in a spin saturated system, 
Eqs. (\ref{matrixelement1}) and (\ref{matrixelement2}) below must be 
multiplied by a factor $1/s_\rho(s_\rho+1)=1/3$, as in
Eq.~(\ref{rholossrate}). Since in Eq.~(\ref{pilossrate}) this factor is  
missing, this rate contains already the sum over all final $\rho$ spin 
states . 

The dependence of the matrix element on the pion momentum $p_i$ arises 
from the derivative in the $\rho\pi\pi$-coupling (\ref{Hint}), whereas 
the one on $p_3=p_\rho$ comes from the $\rho$-polarization tensor \cite{ItzyksonB}  
which appears after summing over all three spin states. If the meson masses 
for different isospin states are taken to be equal, 
(\ref{matrixelement1}) can be simplified by using the momentum 
conserving $\delta$-functions in (\ref{pilossrate},\ref{rholossrate}) 
to evaluate $p_i \pm p_j$ and exploiting the mass-shell conditions 
$p_i^2=m_i^2$: 
 \begin{equation}
 \label{matrixelement2}
    {\cal M} (p_1,p_3) = {\cal M} (p_2,p_3) = 
    \frac{1}{4}\left( m_\rho^2 - (2 m_\pi)^2 \right).
\end{equation} 
In this case the two contributions to the loss rate in 
(\ref{pilossrate}) become equal.   

In kinetic theory the entropy current for bosons or fermions 
is given by
 \begin{eqnarray} 
 \label{entropiegl}
    S_{\mu}(x) 
  &=& -\sum_{i} \int Dp\ p_{\mu} 
    \Bigl[ f_{i} (x,p) \ln f_{i} (x,p)
    \mp ( 1 \pm f_{i} (x,p) ) \ln( 1 \pm f_{i} (x,p) )
  \nonumber\\
  & & \qquad \qquad  \qquad \quad 
    + \bar f_{i} (x,p) \ln \bar f_{i} (x,p) \mp 
   ( 1 \pm \bar f_{i}(x,p) ) \ln( 1 \pm \bar f_{i} (x,p) ) \Bigr],
 \end{eqnarray}
with the upper (lower) sign for bosons (fermions). The rate of 
entropy production can then be written as 
 \begin{equation} 
 \label{entropieprogl}
   \partial^{\mu} S_{\mu} = -\sum_{i} \int Dp  
   \left[ \left( p_{\mu} \partial^{\mu} f_i (x,p) \right)
          \ln\left(\frac{f_i (x,p)}{ 1 \pm f_i (x,p) } \right) 
        + \left( p_{\mu} \partial^{\mu} \bar f_i (x,p) \right) 
          \ln\left(\frac{\bar f_i (x,p)}{ 1 \pm \bar f_i (x,p) } \right)
   \right] .
 \end{equation}
In our $\pi$-$\rho$-system all particles are bosons, i.e. the plus 
signs apply and there are no anti-particle contributions. The sum over 
$i$ in (\ref{entropieprogl}) goes over the three charge states of the 
pions and $\rho$ mesons and additionally over the three spin 
projections of the latter. After replacing $p_{\mu}\partial^{\mu} 
f(x,p)$ in (\ref{entropieprogl}) by the r.h.s. of the kinetic equation 
(\ref{kinequ}), the sum over the three $\rho$ spin states cancels the 
factor $1/s_\rho(s_\rho+1)$ in Eq.~(\ref{rholossrate}), and one gets, 
after some renaming of integration variables and permutation of the 
isospin indices $\alpha, \beta, \gamma$, 
 \begin{eqnarray} 
 \label{entropieproduktion}
    \partial^{\mu} S_{\mu} 
  &=& (2 \pi)^{4} f_{\rho\pi\pi}^{2} \sum_{\alpha \beta \gamma} 
    \vert \varepsilon_{\alpha \beta \gamma} \vert
    \int Dp_1 \, Dp_2 \, Dp_3\, {\cal M} (p_1,p_3)\, 
    \delta(p_1 + p_2 - p_3)
  \nonumber\\
  & & \times 
    \Bigl( 1 + f_\alpha(p_1) \Bigr) \Bigl( 1 + f_\beta(p_2) \Bigr)
    F_{-\gamma}(p_3)\, (y - 1) \ln y \, ,
 \end{eqnarray}
where  
 \begin{equation}
 \label{yequation}
   y = \frac{f_\alpha(p_1) f_\beta(p_2) 
             \left( 1 + F_{-\gamma}(p_3) \right)}
            {\left(1+f_\alpha(p_1)\right) \left(1+f_\beta(p_2)\right)
             F_{-\gamma}(p_3) }\, .
 \end{equation}
$y$ and $\cal M$ are positive by definition. Since for $y \ge 0$ the 
expression $(y - 1) \ln y$ is positive semidefinite, we have 
$\partial^{\mu} S_{\mu} \ge 0$ always. This proves the Boltzmann 
H-theorem in our case and for similar systems with symmetric isospin 
structure. The collision integral (\ref{entropieproduktion}) vanishes 
only if $y = 1$.  

Expressing (similarly to Eq.~(\ref{entropiegl})) the isospin current 
and the energy momentum tensor through the distribution functions $f_i 
(x,p)$ \cite{Heinz85}, one can use the conservation laws for these 
quantities together with the condition for vanishing entropy 
production, $y = 1$, to derive the local equilibrium Bose distribution 
 \begin{equation}
 \label{eqdistr}
   \frac{f_i (x,p_i)}{ 1 + f_i (x,p_i)} =
   \exp \Bigl( -\left[\mu (x) q_i + u_{\nu}(x) p_i^{\nu}\right]/T(x) 
        \Bigr)
   \qquad {\rm or} \qquad
   f_i(x,p_i) = {1\over 
   e^{\left[\mu (x) q_i + u_{\nu}(x) p_i^{\nu}\right]/T(x)} - 1}\, ,
 \end{equation}
where $q_i=0,\pm 1$ is the electric charge of species $i$.

Important for a simulation of the kinetic equations will be the decay 
rate of the $\rho$ meson. In the rest frame of a $\rho$ with a given 
spin polarization it can be cast in the form 
 \begin{equation}
   \label{zerfallsrate}
   \Gamma_{\rm loss}^{\rho^\alpha}(m_\rho) = 
   \underbrace{
   \frac{f_{\rho\pi\pi}^2}{ 16 \pi s_\rho(s_\rho+1)} 
   \frac{\left( m_\rho^2 - (2 m_\pi)^2\right)^{3/2}}{m_\rho}
   }_{\mbox{free decay rate $\Gamma_{\rm free}$}}\cdot
   \frac{\vert \varepsilon_{\alpha\beta\gamma} \vert}{2}
   \int d(\cos\theta)
   \left( 1 + f_\beta (\vec p) \right)
   \left( 1 + f_\gamma (-\vec p) \right).
 \end{equation}
The prefactor is the free decay rate $\Gamma_{\rm free}$, and the 
integral over the pion  
emission angles relative to the velocity of the $\rho$ in the local 
heat bath gives the medium corrections due to Bose stimulated decay.  
The pion distribution functions under the integral are evaluated at
$\vert p \vert = \sqrt{ m_{\rho}^2 - 4 m_{\pi}^2}$, the pion momentum
in the rest frame of the decaying $\rho$ meson.
 
In the limit of free decay in the vacuum the kinetic equation for 
$\rho$ mesons takes the following simple form 
 \begin{equation}
   E_{\rho} \, \frac{d}{dt} F_\alpha (p_\rho,t) =
   - \Gamma_{\rm free} F_\alpha (p_\rho,t) \, .
\end{equation}
This is solved by 
 \begin{equation}
   F_\alpha (p_\rho,t) = F_\alpha(p_\rho,t_0) 
   e^{- \left(\Gamma_{\rm free}/E_\rho\right) (t - t_0)}.
\end{equation}
The measured decay rate in the rest frame of the $\rho$ meson 
($E_{\rho} = m_{\rho}$) is then given by 
 \begin{equation}
 \label{freiezerfallsrate}
   \Gamma_\rho^{\rm decay}(m_\rho^2) = \frac{f_{\rho\pi\pi}^2}{48 \pi}
   \frac{(m_\rho^2 - 4 m_\pi^2)^{3/2}}{m_\rho^2}. 
 \end{equation}
Its experimental value \cite{databook} at the maximum of the resonance 
at ${\overline{m}}_{\rho} = 769.9 \pm 0.8$ MeV is $151.2\ \pm 1.2\ 
\mbox{MeV}$. Together with the measured value of $m_{\pi^\pm} = 
139.56995 \pm 0.00035$ MeV \cite{databook} this determines the 
coupling constant as $f_{\rho \pi \pi} = 6.049 \pm 0.027$.  

For the simulation of the collision term we need a collision criterium 
which involves the $\pi \pi$-scattering cross section $\pi(p_1) + 
\pi(p_2) \rightarrow \rho(p_\rho)$. Assuming a Breit-Wigner mass 
distribution for the $\rho$ meson 
 \begin{equation}
 \label{BreitWigner}
   W(m_\rho^2) = {1\over \pi}\,
                 {\overline{m}_\rho\Gamma(m_\rho^2) \over
                  (\overline{m}_\rho^2 - m_\rho^2)^2 + 
                  \overline{m}_\rho^2 \Gamma^2(m_\rho^2)} \, ,\qquad
                \int_{4 m_\pi^2}^\infty W(m_\rho^2) dm_\rho^2 = 1\, ,
 \end{equation}
with a width given by the free decay 
rate $\Gamma_\rho^{\rm decay}(m_\rho^2)$, Eq.~(\ref{freiezerfallsrate}), one
gets, following standard textbook  
methods \cite{Itzykson,Pilkuhn}, 
 \begin{equation}
 \label{wirkungsquerschnitt}
   \sigma(\mu_{\rho}^2;p_\rho,\gamma) 
   = 4\pi s_\rho(s_\rho+1) f^2 \,
   {\mu_{\rho}\, (\mu_\rho^2 - 4 \mu_\pi^2)^2 
    \over
    (1-\mu_\rho^2)^2 \mu_\rho^4 + f^2 (\mu_\rho^2 - 4 \mu_\pi^2)^3}
   \,  \Bigl( 1 + F_\gamma(p_1 + p_2) \Bigr)
   \cdot \left[\frac{197.327\,{\rm MeV}}{\overline{m}_\rho} \right]^2 
   {\rm fm}^2
 \end{equation}
where 
 \begin{equation}
    f = \frac{f_{\rho\pi\pi}^2}{16 \pi s_\rho(s_\rho+1)}, \quad 
    \mu_\rho = \frac{m_\rho}{\overline{m}_\rho}, \quad
    \mu_\pi = \frac{m_\pi}{\overline{m}_\rho}.
 \end{equation}
Again we have summed over all three spin states of the $\rho$, and the 
factor $( 1 + F_\gamma(p_1 + p_2))$ 
in (\ref{wirkungsquerschnitt}) gives the Bose stimulation  
effect of the medium on $\rho$ production. In an isospin symmetric 
system with $F_+=F_-=F_0$ it does not depend on the final isospin 
state $\gamma$, but only on the momentum $p_\rho = p_1+p_2$ of the 
$\rho$.  
 
\begin{figure}[htb]
 \unitlength1cm
 \begin{center}
 \begin{picture}(9.0,7.5)
 \put(-1.4,-1.2) {\epsfbox{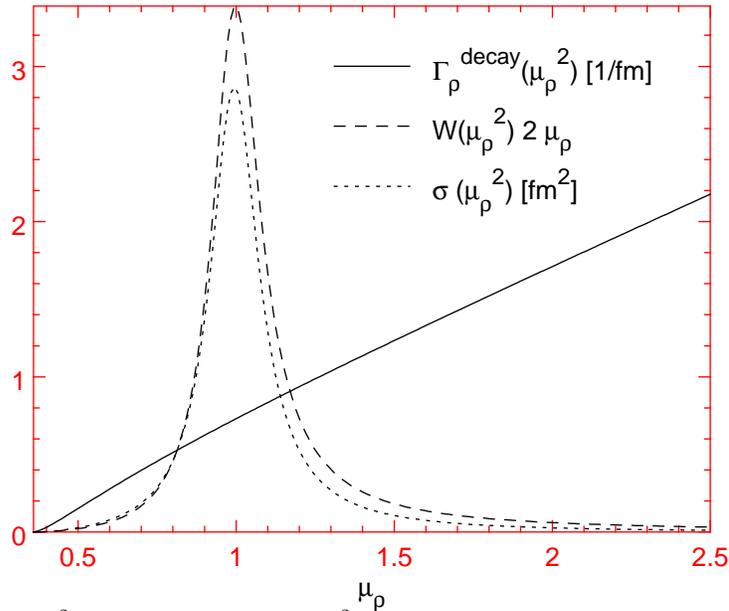}}
 \end{picture}
 \end{center}
 \caption{Free decay rate $\Gamma_\rho^{\rm decay}(\mu_\rho^2)$, free cross 
          section $\sigma(\mu_\rho^2)$, and Breit-Wigner mass 
          distribution $w(\mu_\rho) = 2 \mu_\rho W(\mu_\rho^2)$ for the 
          $\rho$ mesons as functions of their mass in units of
          $\overline{m}_{\rho}$ ,
          the mass at the maximum of the resonance.
          The 
          position of the maxima of $\sigma(\mu_\rho)$ and 
          $w(\mu_\rho)$ are (almost) identical, but at higher masses 
          $\sigma$ has a lower weight than $w(\mu_\rho)$.
 } 
 \label{sigmgampic}
\end{figure}

The cross section and decay rate in vacuum and the mass distribution 
are shown in Figure \ref{sigmgampic} as functions of $\mu_\rho$.  

Finally we want to note that from a formal integration of the kinetic 
equations over the time parameter $\tau$ it can be shown that a system 
with spherical symmetry in $\vec x$ and $\vec p$ remains spherically 
symmetric for all times (see Appendix \ref{appendixA}).

\section{Simulation of the kinetic equations}
\label{sim}

As initial conditions for the solution of the kinetic equations we 
start with a system of $\rho$'s and $\pi$'s in global thermodynamic 
equilibrium at vanishing chemical potential. The thermodynamic 
conditions will be chosen near the expected ``freeze-out" point. The 
particles are distributed homogeneously and isotropically in a 
spherical box in coordinate space with an equilibrium momentum 
distribution $(\exp(E/T) - 1)^{-1}$. The masses of the $\rho$'s 
are distributed with the Breit-Wigner distribution Eq.~(\ref{BreitWigner}). 
For numerical reasons we will employ a cutoff for the $\rho$ mass and 
for the momenta.  

We assume isospin symmetry in the initial state, and since this 
symmetry is conserved by the strong interactions, we can exploit it
to reduce the number of coupled kinetic equations to be solved.
We can sum the three equations for the pion distribution functions 
$f_\alpha$ and the nine equations for the $\rho$ meson distribution 
functions $F_{\alpha s}$, where $\alpha=+,-,0$ and $s=+,-,0$ denote 
isospin and spin projections, respectively. Setting $f_+ = f_- =f_0 \equiv f$ 
and $F_{++} = F_{+-} = \dots \equiv F$, i.e. the distributions for all 
isospin and spin components equal to each other, we find the two 
coupled equations
 \begin{eqnarray}
 \label{pikineq}
   p^\mu \partial_\mu f(x,p) 
   &=& - \Gamma^\pi_{\rm loss} (x,p)\, f(x,p)
     + \Gamma^\pi_{\rm gain} (x,p)\, \bigl( 1+f(x,p) \bigr)\, ,
 \\
 \label{rhokineq}
   p^\mu \partial_\mu F(x,p,m_\rho) 
   &=& - \Gamma^\rho_{\rm loss} (x,p,m_\rho)\, F(x,p,m_\rho)
     + \Gamma^\rho_{\rm gain} (x,p,m_\rho)\, \bigl( 1+F(x,p,m_\rho) \bigr)\, .
 \end{eqnarray}
In the global coordinate system all distribution functions and rates 
are isotropic in momentum space and thus only functions of $E$. The 
rates are given by 
 \begin{eqnarray}
 \label{piloss}
  \Gamma^\pi_{\rm loss} (E)
  &=& 2 {f^2_{\rho\pi\pi} \over 4\pi}
      \int_{2m_\pi}^\infty dm_\rho\, w(m_\rho) {{\cal M}\over \sqrt{E^2-m_\pi^2}}
      \int_{E_-}^{E_+} dE' \,
      f(E') \bigl( 1 + F(E+E',m_\rho) \bigr)\, ,
  \\
 \label{pigain}
  \Gamma^\pi_{\rm gain} (E)
  &=& 2 {f^2_{\rho\pi\pi} \over 4\pi}
      \int_{2m_\pi}^\infty dm_\rho\, w(m_\rho) {{\cal M}\over \sqrt{E^2-m_\pi^2}}
      \int_{E_-}^{E_+} dE'\, 
      \bigl(1+f(E')\bigr) F(E+E',m_\rho)\, ,
  \\
 \label{rholoss}
  \Gamma^\rho_{\rm loss} (E,m_\rho)
  &=& {1\over 3} {f^2_{\rho\pi\pi} \over 4\pi}
      {{\cal M}\over \sqrt{E^2-m_\rho^2}} 
      \int_{E^-}^{E^+} dE'\, 
      \bigl( 1 + f(E') \bigr) \bigl( 1 + f(E-E') \bigr)\, ,
  \\
 \label{rhogain}
  \Gamma^\rho_{\rm gain} (E,m_\rho)
  &=& {1\over 3} {f^2_{\rho\pi\pi} \over 4\pi}
      {{\cal M}\over \sqrt{E^2-m_\rho^2}}
      \int_{E^-}^{E^+} dE'\,
      f(E') f(E-E')\, ,
 \end{eqnarray}
with ${\cal M}$ being the spin-summed matrix element 
(\ref{matrixelement2}). The integration limits are
 \begin{eqnarray}
 \label{limits}
  E_\pm &=& {1\over 2m_\pi^2} \left( E(m_\rho^2-2m_\pi^2) \pm 
                        p\, m_\rho\sqrt{m_\rho^2-4m_\pi^2} 
                 \right)\, , 
  \nonumber\\
  E^\pm &=& {1\over 2} \left( E \pm 
                        {p\over m_\rho}\sqrt{m_\rho^2-4m_\pi^2} 
                 \right)\, , 
 \end{eqnarray}
and $w(m_\rho) = 2 m_\rho W(m_\rho)$ is the mass distribution, 
Eq.~(\ref{BreitWigner}), for the $\rho$ mesons.  
The distribution functions are normalized according to 
\begin{eqnarray}
  \label{normpi}
  N_\pi = N_{\pi^+} + N_{\pi^-} + N_{\pi^0} 
  &=& 3 \int {d^3r\ d^3p \over (2\pi)^3} f(\vec r, \vec p, t)
  \nonumber\\
  &=& 3 {2\over \pi} \int r^2dr \int dE\ E\,\sqrt{E^2-m_\pi^2}
  f(r,E,t) \, ,
  \\
  \label{normrho}
  N_\rho = N_{\rho^+} + N_{\rho^-} + N_{\rho^0} 
  &=& 9 \int {d^3r\ d^3p \over (2\pi)^3} \int
  dm_{\rho}\ w(m_{\rho}) 
  F(\vec r, \vec p, m_{\rho}, t) 
  \nonumber\\
  &=& 9 {2\over \pi} \int r^2dr 
  \, \int_{2 m_\pi}^{\infty} dm_{\rho}\  w(m_{\rho}) 
  \int_{m_\rho}^{\infty} dE E\sqrt{E^2-m_\rho^2}
  F(r,E,m_{\rho},t) \, ,
\end{eqnarray}
with $N_\rho+N_\pi/2 = N$ being constant during the time evolution.

In the entropy production rate, Eq.~(\ref{entropieproduktion}), all 6 
terms of the sum become equal, and it simplifies to 
 \begin{eqnarray}
 \label{entprod}
   \partial^\mu S_\mu 
   &=& 6 (2\pi)^4 f^2_{\rho\pi\pi}
       \int  dm_{\rho}\,w(m_{\rho})\,{\cal M}(m_\rho)\int Dp_1\, Dp_2\, Dp_3\, 
       \delta(p_1+p_2-p_3)\, 
   \nonumber\\
   && \qquad \times
   \ln \left( {f(p_1) f(p_2) \bigl(1+F(p_3,m_{\rho})\bigr) \over
               \bigl(1+f(p_1)\bigr) \bigl(1+f(p_2)\bigr) F(p_3,m_{\rho})}
       \right)
   \nonumber\\
   && \qquad \times
   \Bigl( f(p_1) f(p_2) \bigl(1+F(p_3,m_{\rho})\bigr)  
        - \bigl(1+f(p_1)\bigr) \bigl(1+f(p_2)\bigr) F(p_3,m_{\rho}) \Bigr) \, .
 \end{eqnarray}
With the help of the four-dimensional $\delta$-function the $p_3$ and
$m_\rho$ integrations can be evaluated. This leads to
\begin{eqnarray}
  \label{entprod2}
  \partial^\mu S_\mu 
  &=& 6 \pi f^2_{\rho\pi\pi} \int Dp_1\, Dp_2\, 
  w(m_{\rho})\,\frac{
  {\cal M}(m_\rho)
  }
  {m_{\rho}} 
  \nonumber\\
  && \qquad \times
  \ln 
  \left(
    {f(p_1) f(p_2) \bigl(1+F(p_3,m_{\rho})\bigr) \over
      \bigl(1+f(p_1)\bigr) \bigl(1+f(p_2)\bigr) F(p_3,m_{\rho})}
  \right)
  \nonumber\\
  && \qquad \times
  \Bigl( f(p_1) f(p_2) \bigl(1+F(p_3,m_{\rho})\bigr)  
  - \bigl(1+f(p_1)\bigr) \bigl(1+f(p_2)\bigr) F(p_3,m_{\rho}) \Bigr) \, ,\\
  && \mbox{with} \quad     \vec{p}_3 = \vec{p}_1 + \vec{p}_2,
  \qquad m_{\rho} = \sqrt{
    (E_1 + E_2)^2 - (\vec{p}_1 + \vec{p}_2)^2}. \nonumber
\end{eqnarray}

To simulate the time evolution of the particles we wrote a cascade 
code in which the distribution functions $f$ and $F$ are represented 
by isospin symmetric test particles. In terms of the initial radius $R$ 
of the system the conserved particle number $N$ is given by $N = 
\varrho \cdot R^3$, where $\varrho$ involves the initial $\pi$ and 
$\rho$ densities and is a function of temperature, see Eqs.~(\ref{normpi},
\ref{normrho}). The initial radius of the fireball was fixed to be $5$ fm;
the corresponding particle numbers for temperatures $T$ = 100, 150 and
200 MeV are given in Table~\ref{tabgleichgewicht}.  

 \begin{table}[ht]
 \caption{Particle numbers, density $\varrho$, and the 
          conserved particle number $N$ for a system of $\rho$ mesons
          and pions in equilibrium at $T$ = 100, 150, 200 MeV, for 
          a fireball radius of $R = 5$ fm.
         }
 \label{tabgleichgewicht}
  \begin{center} 
    \begin{tabular}{|c||c|r|l|r|}
      \hline \phantom{\Big\vert}
      $T$ [MeV] & $N_\pi$ 
      & $N_\rho$ & $\varrho [\mbox{fm}^{-3}]$ & $N = N_{\pi}/2 + N_\rho$ \\
      \hline \hline
      100 \hfill&\hfil 16.8 \hfil&\hfil 0.6 \hfil&\hfil 0.07 \hfil&
      \hfil 9 \hfil \\
      \hline
      150 \hfill&\hfil 61.8 \hfil&\hfil 11.1 \hfil&\hfil 0.34 \hfil&
      \hfil 42 \hfil \\
      \hline
      200 \hfill&\hfil 172.0 \hfil&\hfil 70.0 \hfil&\hfil 1.25 \hfil&
      \hfil 156 \hfil \\
      \hline
    \end{tabular}
  \end{center}
 \end{table}

\begin{figure}[htb] \unitlength1cm 
 \begin{center} 
 \begin{picture}(9.0,7.5) 
  \put(-1.4,-1.2) {\epsfbox{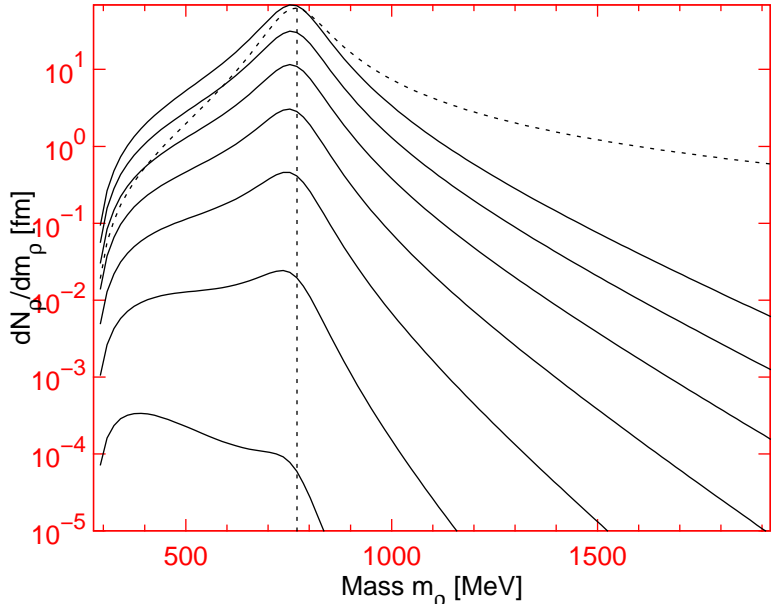}} 
 \end{picture}  
 \end{center} 
 \caption{Mass distributions for $\rho$ mesons in equilibrium with pions for
   different temperatures. The temperatures from top to bottom are $T$ =
   200, 175, 150, 125, 100, 75, and 50 MeV. Also shown for comparison
   is a pure Breit-Wigner mass distribution (dotted line), normalized
   at its peak to the $T$ = 200 MeV curve.
         } 
 \label{mdistpic}
 \end{figure}

For the evaluation of the integrals over the $\rho$ mass distribution
a cutoff at $2.5\ \overline{m}_\rho = 1920$ MeV was introduced, and 
the momentum integrals were cut off at $p_{\rm max} =  3.5 
\overline{m}_\rho = 2689$ MeV. The mass cutoff was large enough to have no 
influence on our results in spite of the fact that for this cutoff the 
tail of the Breit-Wigner distribution still contains about 10\% 
of the mass spectrum. The reason for this is the exponential form of 
the bosonic energy distribution in Eq.~(\ref{normrho}) which suppresses 
the formation of high-mass $\rho$'s and results in a deviation of the 
mass distribution for thermal $\rho$'s from the mass distribution of 
$\rho$'s at rest (see Fig.~\ref{mdistpic}). Thereby the peak of the 
mass spectrum and its weight is shifted towards lower masses as 
the temperature decreases. A second peak close to the threshold of 
$\rho$ production from two pions appears for temperatures below T 
$\approx$ 75 MeV. It dominates the spectrum for low temperatures,
but since we take the $\pi$'s and $\rho$'s to be in local equilibrium 
the number of $\rho$ mesons becomes very small for these temperatures 
and vanishes for $T$ = 0.

The code determines simultaneously the space and momentum coordinates 
of all test particles as functions of a global time parameter on a 
hyper-surface in phase space. (For an introduction to (nuclear) cascade 
codes see \cite{dasGupta},\cite{Wolf90}; the problem of a simultaneous 
relativistic time parameter is discussed for instance in 
\cite{Kodama84}, \cite{Sorge89}.) Since no mean field effects are 
taken into account, the particles propagate freely on straight lines 
between $\rho$ formation and decay: 
 \begin{eqnarray}
   \vec{x}(\tau + \Delta \tau) 
   & = & \vec{x}(\tau) + \vec{\beta} \Delta \tau,
         \qquad  \vec\beta = \frac{\vec{p}}{E}  \, ,
   \nonumber\\
   \vec{p}(\tau + \Delta \tau) 
   & = & \vec{p}(\tau)\, .
 \end{eqnarray}
$\tau$ is the global time parameter.

The formation of a $\rho$ meson in the scattering of two pions is 
implemented via the collision criterion 
 \begin{equation}
 \label{criterium}
   \pi b^2 \le 2\, \sigma(m_\rho^2,E_\rho)\, .
 \end{equation}
The factor 2 arises from the same factor in front of the matrix 
element in Eqs.~(\ref{piloss},\ref{pigain}). $\sigma(m_\rho^2, 
E_\rho)$ is the medium-modified cross section (\ref{wirkungsquerschnitt}), 
and $b^2$ is the relativistically invariant impact parameter \cite{Sorge89} 
 \begin{equation}
 \label{abstandcm}
   b^2 = - \left(\Delta x - \frac{\Delta x\cdot P}{P^2} P \right)^{\!2} 
       = - (\Delta x)^2 + \frac{(\Delta x \cdot P)^2}{P^2}.
 \end{equation}
Here $P = p_1 + p_2$ is the momentum of the pion pair, and $P^2 = 
m_\rho^2$ is the square of their invariant mass. In the center-of-mass 
system of the two pions the impact parameter becomes $ b^2 = {\Delta 
\vec{x}^2_{\rm cm}}$. If $b$ satisfies the collision criterion 
Eq.~(\ref{criterium}) the collision of two pions takes place in the global 
system at the point where their distance 
 \begin{equation}
   - (\Delta x)^2 = (\Delta \vec x)^2 = 
   \Bigl(\Delta \vec x(\tau_0) + \Delta \vec \beta \cdot \tau\Bigr)^2,
   \qquad \Delta \vec{\beta} = 
   \frac{{\vec p}_1}{E_1} - \frac{{\vec p}_2}{E_2}
 \end{equation}
is minimal. This happens at the collision time $\tau_c$ which is 
given by 
 \begin{equation}
   \tau_{c} = - \frac{\Delta \vec x (\tau_0){\cdot}\Delta \vec \beta}
   {(\Delta \beta)^2},
 \end{equation}
and which is measured relative to the last grid point $\tau_0$ of the 
global time parameter at which all trajectories were calculated.  

For the decay of a $\rho$ meson we assume isotropic decay in its
rest frame. To decide if a decay in the rest frame of the $\rho$ 
takes place we use the mass-dependent decay rate Eq.~(\ref{zerfallsrate})
which involves the pion distribution function in the rest frame of the
decaying $\rho$ meson. We evaluate the latter by counting pions 
in phase space cells around the pion momentum $\vec{p}$, $\vert 
\vec{p} \vert = \sqrt{m_\rho^2 - 4  m_\pi^2}$. To avoid the transformation 
of all pion coordinates from the global system into the rest frame of 
the decaying $\rho$ meson, we transform instead the phase space volume 
into the global system and do the counting there.  

The phase space contributions to the cross section and 
decay rate are averaged over a large number of parallel ensembles. The latter 
influence each other only via the evaluation of the Bose factors as an 
ensemble average. To obtain from the discrete locations of the test 
particles smooth, but positive definite phase space density 
distributions, we smear out the contribution from a particle at some 
point $(x,p)$ over phase space with a Gaussian weight function.  
According to the uncertainty principle the smallest possible phase 
space cell has the volume $\Delta^3 x \Delta^3p = h^3$, thus the
widths of the weight functions in $x$ and $p$ have to  
fulfill $\Delta x \Delta p = 2 \pi \hbar $.  
 
In evaluating the collision term, including the Bose enhancement 
factors, with probabilistic methods, we use a trick similar to the one 
described in \cite{WelkeApril91} (see also 
\cite{WelkeMay91,WelkeFeb92}). Instead of computing for each possible 
$\rho$ decay or creation the exact contributions from the Bose factors 
to the decay rate and cross section, we first determine the decay or 
formation probability using an upper bound for the Bose factors (and 
thus the transition rate). Only if the collision would actually happen 
under these conditions, we proceed to check the collision criterium 
with a more realistic estimate of the Bose enhancement factors. This 
saves an appreciable amount of computer time by eliminating without 
effort a large number of ``unsuccessful collision attempts". The upper 
bound on the Bose factors is refined in two successive steps: In a 
first step we obtain a global (and time-independent) upper limit on 
the Bose factors from the initial phase-space density of the expanding system. 
If by using this global limit we find that $\rho$ formation or decay 
are possible, we calculate a more accurate dynamical bound from the 
actual density distributions of the previous time step.  
Only if this second upper bound also allows $\rho$ formation and decay
do we actually calculate the distribution functions as described above.

In our simulations the number of particles in each system, $N = N_{\pi}/2 
+ N_\rho$, times the number of parallel ensembles %$N_{\mbox{para}}$
was fixed to $\approx$ 300000. We followed the system for 150 time 
steps of $\Delta \tau = 0.1$ fm/c.

\section{Results}
\label{res}

\subsection{Decay and re-scattering rates}
\label{decsec}

In general the decay of the $\rho$ mesons is governed by the decay law 
 \begin{equation}
 \label{decaylaw0}
  N_{\rho}(\tau + \Delta \tau) =  
  {6 \over \pi} R^3 \int_{2 m_\pi}^{m_{\rm max}} dm_{\rho}\,
  w(m_{\rho}) \int_0^{p_{\rm max}}  dp\ p^2 \, F(E,m_\rho,\tau)
  e^{- \Gamma^\rho_{\rm loss}(E,m_\rho)\Delta\tau/E},
\end{equation}
where $E^2 = p^2 + m_\rho^2$ and $\Gamma^\rho_{\rm loss}$ is the loss rate
for $\rho$ mesons from Eq.~(\ref{rholoss}). If we neglect the influence 
from Bose stimulation by surrounding pions we can replace
$\Gamma^\rho_{\rm loss}/E$ in the exponent by 
$\Gamma_\rho^{\rm decay}(m_\rho)/\gamma$ from Eq.~(\ref{freiezerfallsrate})
where $\gamma$ is the Lorentz factor between the global system and 
the $\rho$ rest frame. 

\begin{figure}[htb] \unitlength1cm 
 \begin{center} 
 \begin{picture}(9.0,7.5) 
 \put(-1.4,-1.2) {\epsfbox{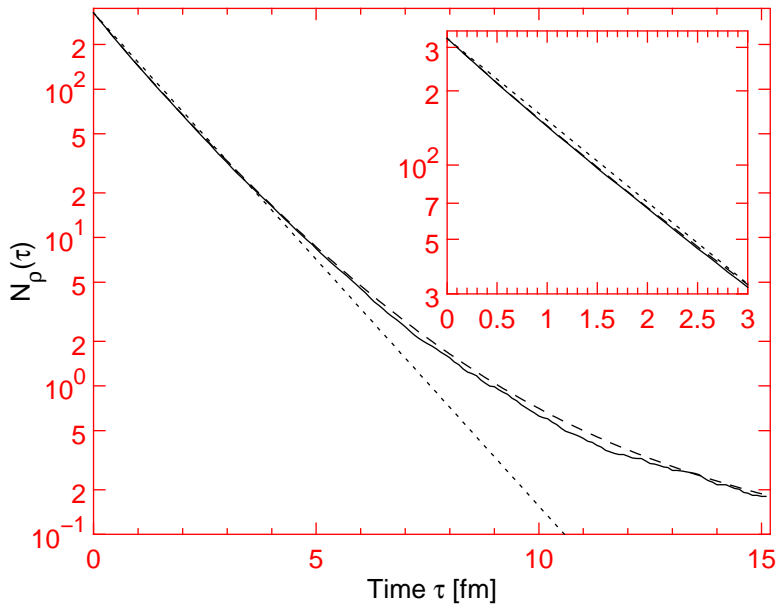}} 
 \end{picture}  
 \end{center} 
 \caption{\label{puredecpic}
          The number of $\rho$ mesons as a function of time for free 
          decay at rest simulated with 1000 parallel ensembles each
          containing initially 300 $\rho$ mesons (solid) and
          calculated according to Eq.~(\protect\ref{decaylaw}) (dashed). 
          For comparison we show the free decay for a 
          constant decay width \protect$\Gamma^{\rm decay}_\rho$ = 151 MeV 
          (dotted).}  
 \end{figure}

For the decay of $\rho$ mesons at rest ($\gamma=1$) into the vacuum one 
then obtains
 \begin{equation}
   \label{decaylaw}
   N_\rho(\tau + \Delta \tau) = \int dm_\rho\,  N_\rho(m_\rho,\tau)\,
   \exp\bigl(-\Gamma_\rho^{\rm decay}(m_\rho) \Delta \tau/\gamma\bigr),
 \end{equation}
where $N_\rho(m_\rho,\tau)$ is the number of $\rho$ mesons with mass 
$m_\rho$ at time $\tau$. Figure \ref{puredecpic} shows a numerical 
simulation of this decay law with the help of our kinetic code. 
There is good agreement between the simulated result and the one 
calculated directly from Eq.~(\ref{decaylaw}). The striking deviation of 
both these results from the straight line describing the exponential 
decay of $\rho$ 
mesons with mass $\overline{m}_\rho = 770$ MeV and a constant width 
$\Gamma^{\rm decay}_\rho=151$ MeV arises from the mass dependent 
decay rate (\ref{freiezerfallsrate}): The heavier $\rho$ mesons from 
the initial Breit-Wigner mass distribution decay first (due to their 
larger than average width the slope of the decay curve is at small 
$\tau$ actually slightly steeper than the reference line), leaving at 
later times only the more long lived lighter $\rho$ mesons from the 
lower end of the mass distribution. Therefore the decay curve levels 
off at large $\tau$.  

\begin{figure}[htb]
\unitlength1cm
\begin{center}
\begin{picture}(9.0,7.5)
 \put(-1.4,-1.2) {\epsfbox{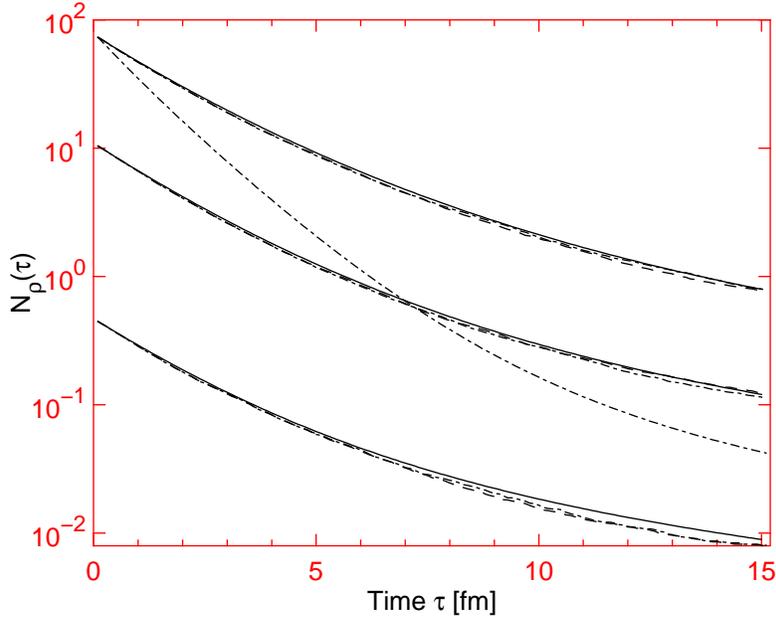}} 
\end{picture}
\end{center}
\caption{\label{rhodecpic} $\rho$ decay for systems without
  $\rho$ formation for different initial temperatures.
  Dashed and dashed-dotted: simulation with and without phase space
  contributions  to the decay rate. Solid: calculated time evolution
  for the decay of $\rho$ mesons with initial equilibrium
  distribution. 
  From top to bottom: $T$ = 200, 150 and 100 MeV.
  For comparison: center of mass decay (dashed-dotted) as calculated with 
  Eq.~(\protect\ref{decaylaw}) and fitted to the initial number of 
  $\rho$ mesons for $T$ = 200 MeV.
  }
\end{figure}

Fig.~\ref{rhodecpic} shows the $\rho$ meson decay for initially 
thermalized systems with different initial temperatures when we neglect 
the back-reaction $\pi\pi\to\rho$. One clearly sees a deviation 
from the just discussed decay of $\rho$ mesons at rest, shown again
by the dotted curve. A comparison between the dashed and dash-dotted 
curves in Fig.~\ref{rhodecpic} shows that this effect is not due to 
medium effects via Bose stimulation from the pions in the system.
It is, however, rather accurately reproduced by an analytical calculation
(solid lines) which uses Eq.~(\ref{decaylaw0}) with a thermal distribution 
function and again neglects Bose stimulation:
 \begin{equation}
 \label{decaylaw2}
  N_{\rho}(\tau + \Delta \tau) =  
  {6 \over \pi} R^3 \int_{2 m_\pi}^{m_{\rm max}} dm_{\rho}
  w(m_{\rho}) \int_0^{p_{\rm max}}  dp\ p^2 
  \left[ e^{\sqrt{p^2 + m_\rho^2}/T} - 1 \right]^{-1} 
  e^{- \Gamma_\rho^{\rm decay}(m_\rho)\Delta\tau/\gamma(p,m_\rho)}.
 \end{equation}

For all three initial temperatures the curves in Fig.~\ref{rhodecpic} 
have approximately the same form. This was not expected, since the 
Lorentz-dilatation factor $\gamma(p,m_\rho) = \sqrt{p^2 + m_\rho^2}/m_\rho$
in the exponent of Eq.~(\ref{decaylaw2}) is weighted quite differently
at different temperatures, which should lead to an effectively larger
decay rate at lower temperature (with smaller average 
$\langle\gamma\rangle$). However, at lower initial temperatures also 
the mass spectrum of the $\rho$ mesons is shifted towards lower masses 
(see Fig.~\ref{mdistpic}). Since the $\rho$ width decreases with 
decreasing mass $m_\rho$, the net effect is a somewhat lower effective 
$\rho$ meson decay rate at lower temperature.

The observed weakness of medium effects from Bose stimulation shows that
the chosen initial conditions select systems which are already 
quite dilute in phase space. Bose statistical corrections would play
a larger role if the systems were not allowed to expand 
\cite{WelkeApril91,WelkeMay91,WelkeFeb92}.

\begin{figure}[htb]
\unitlength1cm
\begin{center}
\begin{picture}(16.0,7.5)
\put(-2.1,-1.2){\epsfbox{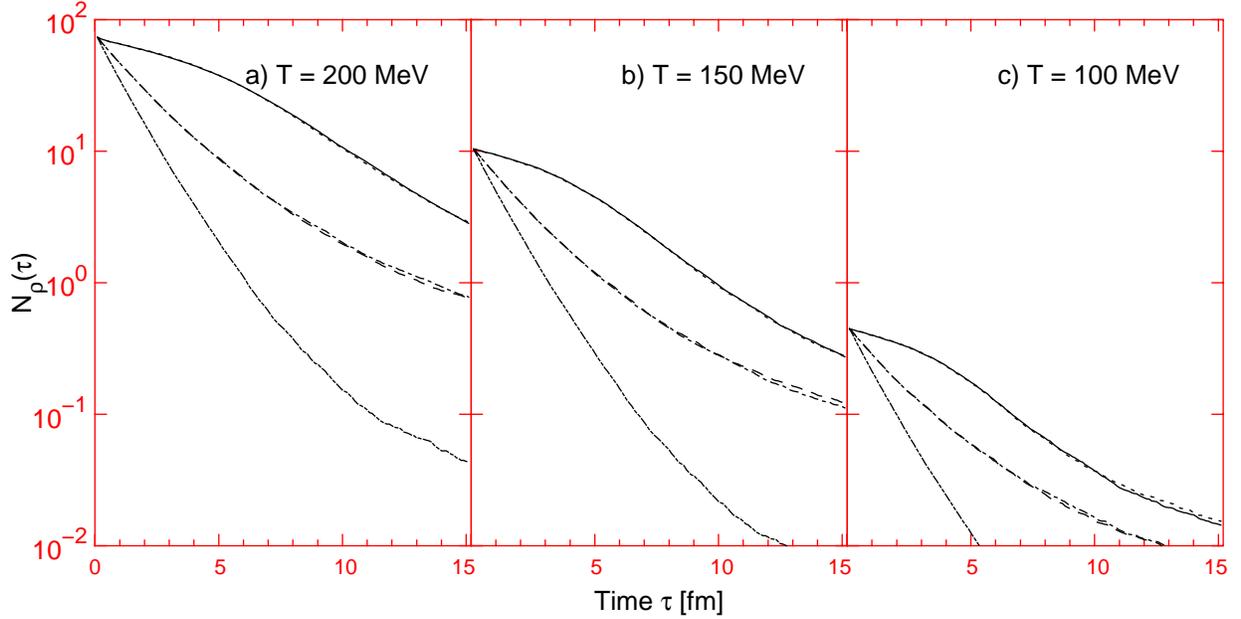}}
\end{picture}
\end{center}
\caption{
         \label{rhototpic} 
         $\rho$ decay from our simulation for systems with different 
         initial temperatures. Curves from top to bottom: 
         $\rho$ decay plus back reaction $\pi\pi\to\rho$, with (solid) and 
         without (dotted) Bose enhancement factors;
         $\rho$ decay without back reaction, with (dashed) and without
         (dash-dotted) Bose enhancement factors;
         free decay rate ($-\,\cdot\,\cdot\,-\,\cdot\,\cdot\,-$)
         for comparison.
         }
\end{figure}

In Fig.~\ref{rhototpic} we study the additional effects from the
back-reaction $\pi \pi \rightarrow \rho$. From a formal point of view 
they must be included in order to preserve detailed balance and the
Boltzmann H-theorem. In practice one sees that the number 
of resonances now decreases much more slowly (solid and dotted curves 
in Figs.~\ref{rhototpic}a-c). Again, the inclusion of Bose enhancement
factors in decay rates and scattering cross sections has no
visible effect.

\begin{figure}[htb]
\unitlength1cm
\begin{center}
\begin{picture}(16.0,7.5)
\put(-2.1,-1.2){\epsfbox{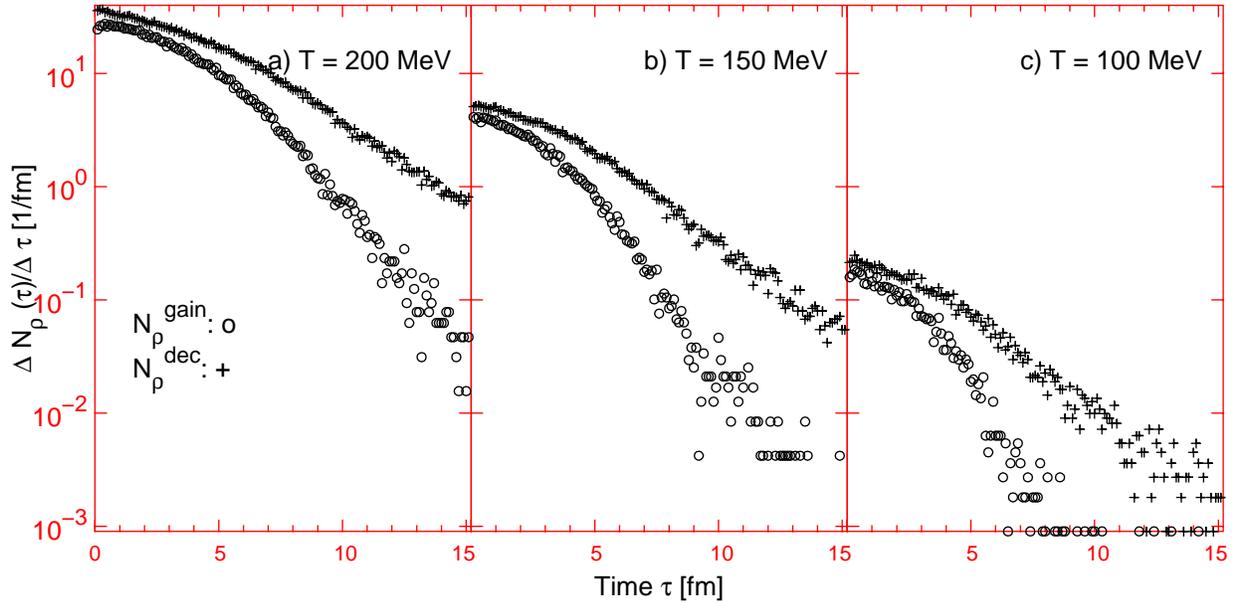}}
\end{picture}
\end{center}
\caption{\label{ratpic} Gain and loss rates of $\rho$ mesons 
         as a function of time for different initial temperatures.}
\end{figure}

Another illustration of the important practical role of $\rho$ 
formation by pion re-scattering is shown in Figs.~\ref{ratpic}
where the total rate for the change of the number of $\rho$ mesons is split
into its loss and gain contributions. Since the system is initialized in
thermodynamic equilibrium, the two contributions should initially balance each
other. That in Fig.~(\ref{ratpic}) they don't quite do that is due to the 
fact that at $\tau=0$ the walls around the initial sphere are removed
and thus $\rho$ mesons from the surface can now escape without
re-scattering. One sees that initially the back reaction rate is still
a considerable fraction of the decay rate, and that
only for large times ($t \geq 5 - 10 $ fm/$c$) the gain of $\rho$
resonances by pion re-scattering can be neglected compared to the losses
by decay. Furthermore, even if the absolute numbers for the decay
and formation rates decrease drastically with decreasing temperature 
due to the exponentially decreasing occupancy numbers, the time dependence
of the ratio between these two rates shows no strong temperature dependence.

In order to better understand these features we give in 
Table~\ref{meanfreepath} the pion mean free path $l^\pi_{\rm free}$ for
different temperatures, as extracted from the mean thermal
velocity $\langle v \rangle_T$ and the effective collision time: 
$l_{\rm free}^\pi = \langle v \rangle_T \tau^\pi_{\rm coll}$.
The mean thermal velocity of the pions at the beginning of the time
evolution is given by
 \begin{equation}
   \langle v \rangle_T = 
   \frac{\int_0^\infty (p/E) [e^{E/T}-1]^{-1} p^2\,dp}
        {\int_0^\infty [e^{E/T}-1]^{-1} p^2\, dp}.
 \end{equation}
The pion (inelastic) collision time can be calculated from $\rho$ 
meson gain rate, using the conservation of $N=N_\rho + N_\pi/2$ 
in the process $\pi+\pi\to\rho$. Defining $\tau^\pi_{\rm coll} = 
- N_{\pi}/\dot N_\pi = + N_\pi/(2\dot N_\rho)$ we get
 \begin{equation}
  \tau^\pi_{\rm coll}(\tau) 
  = \frac{N_\pi(\tau)}
         {2 (\Delta N^\rho_{\rm gain}(\tau)/\Delta \tau)}\, ,
 \end{equation}
with $\Delta N_{\rm gain}^\rho(\tau)$ taken from Figs.~\ref{ratpic} at
$\tau=0$.

The system with $T$ = 100 MeV has a pion mean free path
which is much larger than the initial radius which in our simulations
was chosen as $R=5$ fm. 
If we assume the usual definition of ``freeze out'' which says, that
particles will no longer hit each other if their mean free path
becomes larger than the radius of the system, then ``freeze out'' will
occur immediately after the beginning of the time evolution.
In terms of absolute numbers for $\rho$ formation (Fig.~\ref{ratpic}c) 
this is fulfilled to good accuracy in our simulation.
On the other hand the system with $T$ = 200 MeV has a pion mean free
path which is smaller than the radius of the system, so that, at least
at the beginning of the time evolution, it will stay close to equilibrium.
Nevertheless the shape of the curves for $\rho$ decay in Fig.~\ref{rhototpic} 
and for the differential gain and loss of $\rho$ mesons in Fig.~\ref{ratpic} 
are quite similar for the different temperatures, and
show that the interaction itself does not change its character, only
the number of particles involved gets smaller from $T$ = 200 MeV to $T$ =
100 MeV. Therefore ``freeze out'' is  a quantitative
statement about the number of possible interactions and not 
a qualitative statement about the form of the interaction.

\begin{table}[htb]
\caption{
         Collision time $\tau^\pi_{\rm coll}$, average
         thermal velocity  $\langle v \rangle_T$, and
         mean free path $l^\pi_{\rm free}$ for the pions at the 
         beginning of the kinetic evolution for different
         initial temperatures. The initial radius was 5 fm.
         }
\label{meanfreepath}
  \begin{center} 
    \begin{tabular}{|r||c|r|c|r|}
      \hline
      $ T $ [MeV]
      & $\tau^\pi_{\rm coll}$ [fm/$c$]
      & $\langle v \rangle_T$ 
      & $l_{\rm free}$ [fm] \\
      \hline \hline
      100 & 50.0 & 0.811 & 40.5  \\
      \hline
      150 &  7.6 & 0.867 & 6.6 \\
      \hline
      200 &  3.4 & 0.899 & 3.0 \\
      \hline
    \end{tabular}
  \end{center}
\end{table}

\subsection{Time evolution of the $\rho$ mass spectrum}

\begin{figure}[htb]
\unitlength1cm
\begin{center}
\begin{picture}(16.0,7.5)
\put(-2.1,-1.2) {\epsfbox{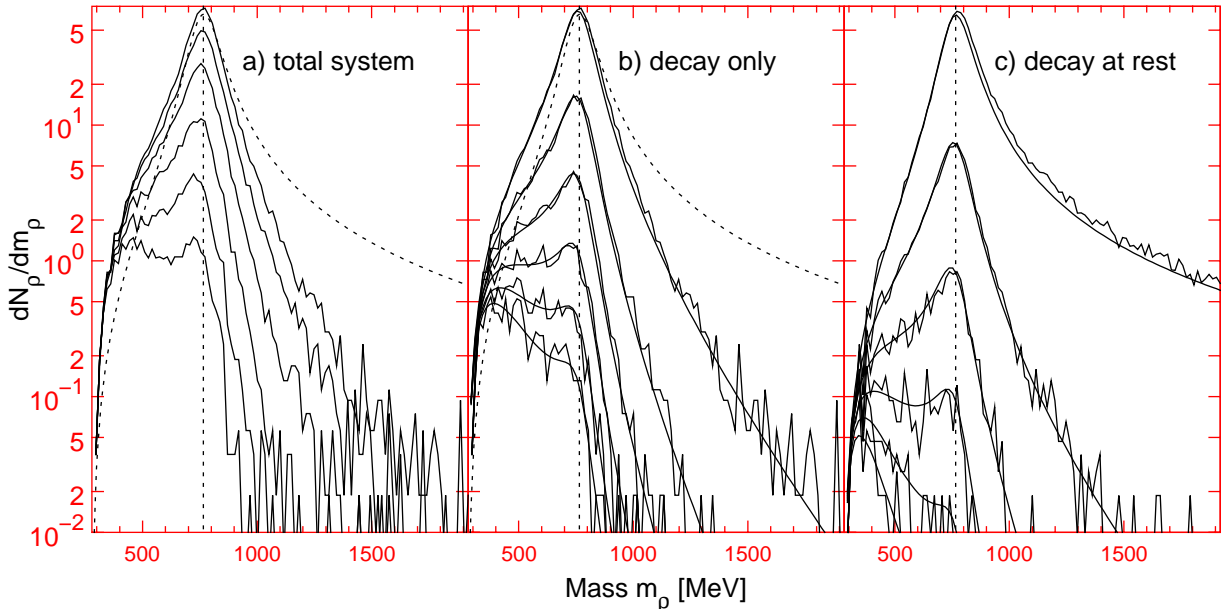}}
\end{picture}
\end{center}
\caption{
         \label{mspeclogpic} 
         Mass spectrum of $\rho$ mesons as a function of time from 
         the kinetic simulation, for an initially thermalized 
         system with temperature $T = 200$ MeV,
         in time steps of 3 fm/$c$. The Breit-Wigner distribution, 
         normalized at its peak (vertical dashed line) to the initial 
         thermally smeared mass distribution, is shown as a dotted line
         for comparison. {\bf a:} total system including decays and
         pion re-scattering. {\bf b:} $\rho$ decays only, with the smooth
         solid curves indicating the result from Eq.~(\protect\ref{decaylaw2}).
         {\bf c:} decay of $\rho$ mesons at rest, with the smooth 
         curves indicating the result from Eq.~(\protect\ref{decaylaw}).
}
\end{figure}

The mass spectrum of the $\rho$ mesons is changed both by 
decay and formation processes, see Fig.~\ref{mspeclogpic}.
For the decay of $\rho$ mesons at rest and for the decay of $\rho$
mesons with an initial thermal momentum distribution we can calculate 
the time evolution for the mass spectrum from Eqs.~(\ref{decaylaw}) 
and (\ref{decaylaw2}), respectively. The results are shown as the
smooth curves in Figs.~\ref{mspeclogpic}b,c. The temporal change of
the mass spectrum arises mostly from the mass dependent decay rate.
We observe that the spectra for large times become qualitatively 
similar in shape to the initial spectra at low temperatures (see 
Fig.~\ref{mdistpic}) and have their peaks shifted towards the 
$\rho$ formation threshold at $2 m_\pi$. Comparing Figs.~\ref{mspeclogpic}b
and \ref{mspeclogpic}a we see that the effect of $\rho$ formation by 
pion re-scattering is a reduction of the effective decay rate for the 
$\rho$ mesons. If in the two different time evolutions we select
points with an equal total number of $\rho$ mesons, the {\em shape} of the 
$\rho$ mass spectrum is, however, nearly identical.

\subsection{Single particle spectra}
 
\begin{figure}[htb]
\unitlength1cm
\begin{center}
\begin{picture}(9.0,7.5)
\put(-1.4,-1.2) {\epsfbox{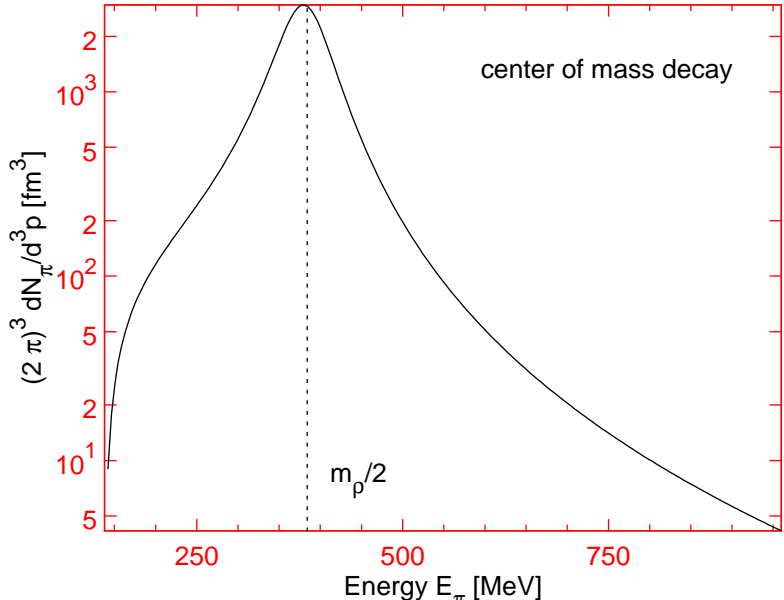}}
\end{picture}
\end{center}
\caption{\label{epspdpic}
         Energy spectrum of pions from decaying $\rho$ mesons at rest.
}
\end{figure}

The pion spectrum from decaying $\rho$ mesons at rest is
fully determined by the initial $\rho$ mass distribution 
and the two-body decay kinematics. This is shown in Fig.~\ref{epspdpic}.
The maximum of the pion energy spectrum at ${E_{\pi}} = 
\overline{m}_{\rho}/2 = 385$ MeV results from the maximum of 
the resonance at $\overline{m}_{\rho} = 770$ MeV,

\begin{figure}[htb]
\unitlength1cm
\begin{center}
\begin{picture}(16.0,7.5)
\put(-2.1,-1.2) {\epsfbox{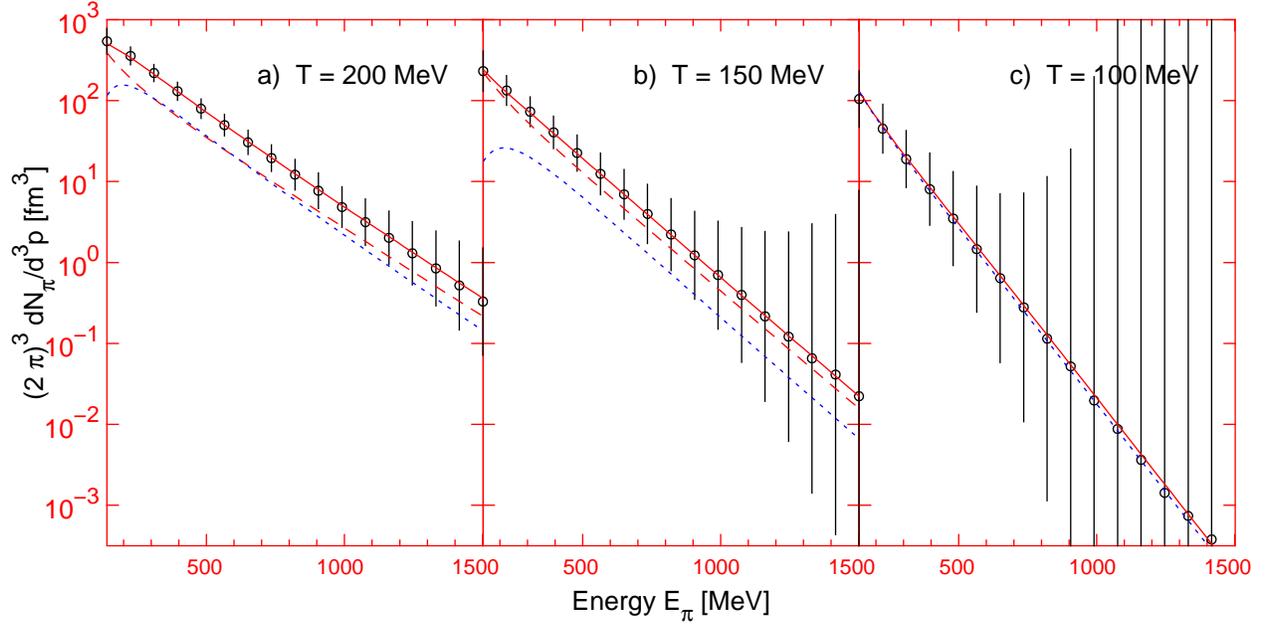}}
\end{picture}
\end{center}
\caption{\label{epstpic} 
         Pion momentum distribution as a function of pion energy
         in the global reference frame at the end of the time evolution.
         The circles indicate the results from the numerical solution,
         which includes both $\rho$ decay and formation by pion
         re-scattering. The error bars indicate the statistical error
         from the sampling over parallel ensembles.
         For comparison: equilibrium (Bose) pion distribution 
         at the beginning of the time-evolution (dashed line), and (as solid
         curve) 
         a superposition of these initial pions with pions from $\rho$
         decays (dotted line), assuming that all initially present $\rho$
         mesons decay 
         directly, without pion re-scattering.}
\end{figure}

In the absence of pion re-scattering, in our model the Lorentz-invariant 
pion momentum spectrum consists of a superposition of the initial 
thermal pion spectrum and the spectrum of decay pions from thermally 
distributed $\rho$ mesons \cite{Sollfrank92a}. This is shown as the 
solid line in Fig.~\ref{epstpic} where the contribution of pions from direct
$\rho$ decay (dotted line) is given by
\begin{eqnarray}\label{directrhoeq}
(2 \pi)^3 \frac{dN_{\pi}}{d^3p} &=& 9\frac{1}{p \sqrt{p^2+m_\pi^2}} \int d^3x
\int\limits_{2 m_\pi}^{\infty} \frac{dm\ m\ w(m)}{\sqrt{m^2 - 4 m_\pi^2}}
\int\limits_{E^-}^{E^+} \frac{E\ dE}{e^{\beta E} -1}\\
&& E^\pm = \frac{m^2}{2 m_\pi^2}\left[ \sqrt{p^2 + m_\pi^2} \pm
  \frac{p}{m}\sqrt{m^2 - 4 m_\pi^2}\right]\nonumber . 
\end{eqnarray}
Once $\rho$ formation by pion re-scattering
is taken into account, one might expect that this changes both the
direct pion spectrum (because pion loss by re-scattering into the $\rho$
channel is not equally distributed in momentum space) and the 
decay spectrum from $\rho$ decays (because the $\rho$'s from re-scattering
pions need no longer have a thermal distribution) after the system is 
allowed to expand. (Of course, initially the decay and re-scattering 
processes don't change either spectrum because the system is in 
thermodynamic equilibrium.) Figs.~\ref{epstpic} show that the continuous 
nature of freeze-out in our kinetic simulation has only very weak effects
on the final energy spectra: even for the rather dense system with initial 
temperature $T=200$ MeV the somewhat steeper final pion spectrum (compared 
to the initial thermal one) is very accurately described by a thermal 
direct pion component superimposed by decay pions from thermally 
distributed $\rho$'s. (We have checked that our simulation, with re-scattering
shut off such that all $\rho$ mesons decay directly with their initial
thermal distribution, agrees with the resonance decay calculations of 
Eq.(\ref{directrhoeq})
.) No additional cooling of the pions by the
re-scattering processes is observed, if it is there it is accurately 
compensated by the developing radial collective expansion. We conclude
that the calculation of the single particle spectra according to the methods
used in Refs.~\cite{Sollfrank92a,Schned92a,Schned93b,Schned94} (which 
do not include any kinetic evolution after the so-called freeze-out point)
provides a quantitatively accurate approximation.

\subsection{Entropy production during expansion}\label{entropysec}

Before attempting to calculate the entropy production
in our system it is very important to gain a rough qualitative picture
of the phase space distributions of the particles. At the beginning 
the distributions in momentum and coordinate space are totally uncorrelated,
due to the assumption of (global) thermal equilibrium. After allowing 
the system to expand, strong correlation between the momenta and coordinates
develop. After a short interactive phase, the system essentially evolves
by free-streaming, and after some time the momentum distribution
at each point in coordinate space becomes essentially a $\delta$-function
in momentum space $\delta(r-(p/E)t)$. This very uneven distribution
in phase-space leads to severe problems if one tries to calculate
the entropy by a naive phase-space integration of $f\ln f$ without 
violating numerically Liouville's theorem that the entropy in system 
of free-streaming particles remains constant.

We approached this problem by a particular binning in phase space 
adapted to a system of free-streaming particles. We evaluated the 
total entropy and the entropy production rate on phase space cells 
of fixed size $\Delta x \cdot \Delta p = 2 \pi \hbar$, whose actual 
shape, however, was adapted to the free expansion by projecting 
the particle coordinates for each time step to the point in space 
where the particles would have started their trajectory if no 
interactions and decays had taken place. We then evaluated the 
particle densities within cubic boxes in the new coordinates, 
thereby obtaining an equivalent ``initial'' phase-space distribution 
for free streaming particles.

This method is applicable as long as only a small additional volume 
in phase space is occupied by particle interactions during time 
evolution. This is the case for systems which freeze out shortly 
after the beginning of the time evolution. If the system
spread out more into the initially available free phase space,
the phase-space densities could become too low for a reliable 
simulation of the kinetic equations by our test particle method.

\begin{figure}[htb]
\unitlength1cm
\begin{center}
\begin{picture}(16.0,7.5)
\put(-2.1,-1.2) {\epsfbox{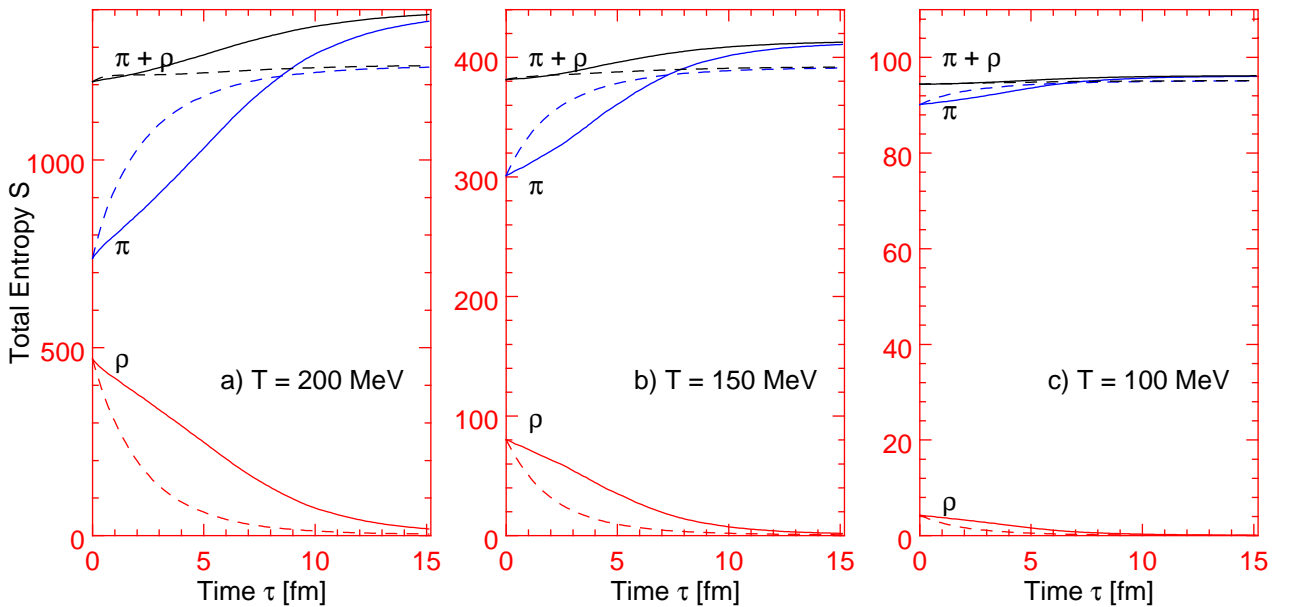}}
\end{picture}
\end{center}
\caption{\label{sigdirpic} 
         The total entropy as well as the individual contributions
         from pions and $\rho$ mesons as a function of time. Dashed curves:
         $\rho$ decay only, no pion re-scattering. Solid curves: Full
         numerical simulation including re-scattering processes.}
\end{figure}

In Fig.~\ref{sigdirpic} we show the calculation of entropy as a function 
of global time $\tau$. It is calculated from Eq.~(\ref{entropiegl}), by 
integrating the entropy current over a hyper-surface of constant global time:
 \begin{equation}
 \label{entint}
   S(\tau) = \int S^\mu(x)\, d\Sigma_\mu^{(\tau)}(x) 
           = \int S^0({\bf x},\tau)\, d^3x \, .
 \end{equation}
Eq.~(\ref{entropiegl}) allows to separate the contributions from pions 
and $\rho$'s. The numerical simulation was stopped after $\tau = 15$ fm/$c$ 
when nearly all $\rho$ mesons had decayed and the entropy had saturated.
Table~\ref{tabentro1} gives the initial equilibrium values for the entropy, 
Table~\ref{tabentro2} the final values and the relative increase during 
the time evolution, both for $\rho$ decays without re-scattering and 
including re-scattering.

\begin{table}[htb]
\caption{\label{tabentro1} 
         Initial equilibrium values for the entropy for different initial
         temperatures.}
  \begin{center} 
     \begin{tabular}{|r||c|c|c|}
       \hline
       T[MeV] 
       & $S_{\rho}$
       & $S_{\pi}$
       & $S_{\rm tot}$ \\
       \hline 
       \hline
       200 & 469.8 & 738.4 & 1208.2 \\  
       \hline       
       150 & 80.3 & 301.1 & 381.5 \\
       \hline
       100 & 4.3 & 90.1 & 94.4 \\
       \hline
    \end{tabular}
  \end{center}
\end{table}

\begin{table}[htb]
\caption{\label{tabentro2} 
         Final entropy and relative entropy increase for three
         different initial conditions. 
         }
   \begin{center}
     \begin{tabular}{|r||c|c|c|c|}
       \hline
       \multicolumn{5}{|c|}{a) $\rho$ decay only} \\
       \hline
       \hline
       T[MeV] 
       & $S_{\rho}$ 
       & $S_{\pi}$
       & $S_{\rm tot}$
       & $\Delta S / S_{\rm tot}$ in \% \\
       \hline \hline 
       200 & 4.2 & 1246.8 & 1251.1 & 4 \\  
       \hline       
       150 & 0.8 & 391.2 & 392.0 & 3 \\
       \hline
       100 & 0.0 & 95.1 & 95.1 & 1 \\
       \hline
       \hline
       \multicolumn{5}{|c|}{b) $\rho$ decay plus pion re-scattering} \\
       \hline
       \hline
       T[MeV] 
       & $S_{\rho}$ 
       & $S_{\pi}$
       & $S_{\rm tot}$
       & $\Delta S / S_{\rm tot}$ in \% \\
       \hline \hline 
       200 & 17.9 & 1369.0 & 1386.9 & 15 \\  
       \hline       
       150 & 1.9 & 410.9 & 412.7 & 8 \\
       \hline
       100 & 0.1 & 96.1 & 96.2 & 2 \\
       \hline
     \end{tabular}
   \end{center}
\end{table}

One sees that $\rho$ decays alone produce very little additional
entropy. Even for an initial temperature of $T$ = 200 MeV, where
at the beginning nearly 30\% of all particles are $\rho$ mesons
(see Table~\ref{tabgleichgewicht}), the total increase in entropy is 
only 4\% of the initial value. Including the re-scattering processes
leads to a somewhat larger entropy increase, by up to 15\% at $T=200$ 
MeV. One sees from the numbers that in general the re-scattering 
processes are much more efficient in producing entropy than
the decays. But even at the highest temperatures investigated in this 
work the entropy increase is modest; for systems which are initialized 
near freeze-out ($T<150$ MeV, see Table~\ref{meanfreepath}) it is well 
below 10\%. This implies that, as far as calculating the entropy 
balance from the final particle distributions is concerned 
\cite{Sollfrank92b}, the fact that freeze-out is not a sudden process 
can be safely ignored.

\section{Conclusions}
\label{con}

We presented a cascade simulation with thermal equilibrium initial 
conditions for the kinetic evolution of the $\pi$-$\rho$-system.  
This work was motivated by an earlier investigation of the influence
of resonance decays on the entropy balance in relativistic heavy-ion 
collisions \cite{Sollfrank92b} which did not take into account that
particle freeze-out is a continuous rather than a sudden process. We have 
answered the question to what extent additional entropy can be created 
during the final kinetic evolution between the time when local thermal 
equilibrium is first broken and the time when finally all resonances 
have decayed. We found that this additional amount of entropy is very
small, less than about 10\% of the entropy already present at the beginning
of decoupling. We also found that this additional entropy production
is dominated by the resonant re-scattering between the pions in the system
rather than by the resonance decay itself. The entropy saturates once
the system is so dilute that re-scattering events become unlikely.

Our simulations took into account the Bose statistics of the mesons,
and we checked that our collision term satisfies Boltzmann's H-theorem
and eventually leads to Bose equilibrium distributions if the matter
is not allowed to expand. We realized, however, that in the calculation
of the collision rates medium corrections by stimulated emission were
very small. Consequently the influence of Bose statistics on the 
kinetics and dynamics of the expanding system turned out to be negligible.

We also found that the final pion single particle spectra resulting 
from the kinetic simulation can be very well approximated by the 
initial thermal pion distribution plus a contribution of pions from 
the decay of $\rho$ mesons with a thermal momentum distribution with
the same initial temperature. In other words, the modifications of the 
spectral shape from the final kinetic evolution of the 
$\pi$-$\rho$-system, in particular the effects of pion re-scattering
into $\rho$-mesons, are minor. Thus the usual method 
\cite{Sollfrank92a,Schned92a,Schned93b} of calculating
the resonance decay contributions to the pion spectrum in the ``sudden 
approximation'', i.e. without considering the detailed kinetic balance
between decays and re-scattering, is a good approximation and
quantitatively quite accurate. This is certainly good news for 
practitioner because it means that for a thermo- and hydrodynamic 
analysis of measured hadron spectra the large numerical efforts of
a detailed kinetic treatment of the freeze-out stage are not necessary.

In the course of our investigations we also discovered a number 
of other interesting aspects of our system. It is well known that
in a thermalized system the $\rho$ mass spectrum is shifted towards
lower masses compared to the vacuum case, due to the weighting
with the exponential thermal Bose distribution (see Eq.~(\ref{normrho})).
However, during the time evolution we observed a further downward 
shift in the mass peak due to the mass dependent decay rate. This 
effect simulates very efficiently an increasingly cooler environment 
for the $\rho$ mesons as time goes on.

\noindent{\bf Acknowlegdement:} 
We thank P. Koch-Steinheimer for stimulating
discussion and constructive advice during the initial stages of this
work. 
U.H. wishes to thank B. M\"uller for stimulating discussions and the Physics
Department at Duke University, where this work was completed, for the warm
hospitality.
This work was supported by DFG, GSI and BMBF.

\begin{appendix}
\section{Appendix}
\label{appendixA}

For a system with spherically symmetric initial conditions in $\vec p$ 
and $\vec x$ we show by formal integration over the global 
time parameter $\tau$ that the kinetic equations conserve the 
symmetry.  

From the kinetic equations we obtain formally at time $\tau + \Delta 
\tau$
 \begin{eqnarray}
 \label{kinequrho}
   F_{\alpha}(\tau+\Delta\tau, m_\rho) = F_{\alpha} (\tau,m_\rho)
  &-& (2\pi)^4 {f_{\rho\pi\pi}^2\over m_\rho s_\rho(s_\rho +1)}\sum_{\beta
  \gamma}\varepsilon_{\alpha \beta \gamma} 
   \int Dp_\beta\, Dp_\gamma\, \delta (p_\beta + p_\gamma - p_\alpha)
   \, {\cal M} 
  \nonumber\\
  & & 
   \Bigl[F_\alpha(\tau,m_\rho) \bigl(1+f_\beta(\tau)\bigr)
                        \bigl(1+f_\gamma(\tau)\bigr) 
       - \bigl(1+F_\alpha(\tau,m_\rho)\bigr)
         f_\beta(\tau) f_\gamma(\tau) 
   \Bigr] 
   \Delta\tau\, ,
 \end{eqnarray}
 \begin{eqnarray}
 \label{kinequpi}
   f_\alpha(\tau+\Delta\tau) =
   f_\alpha(\tau)
  &+& 2 (2\pi)^4 {f_{\rho\pi\pi}^2\over m_\rho}\sum_{\beta
  \gamma}\varepsilon_{\alpha \beta \gamma} 
   \int Dp_\beta \, Dp_\gamma\, \delta (p_\alpha + p_\beta - p_\gamma)
   \, {\cal M}
  \nonumber\\
  & & 
   \Bigl[\bigl(1+f_\alpha(\tau)\bigr)
         \bigl(1+f_\beta(\tau)\bigr) F_\gamma(\tau,m_\rho) 
       - f_\alpha(\tau) f_\beta(\tau) \bigl(1+F_\gamma(\tau,m_\rho)\bigr)
   \Bigr] 
   \Delta\tau \, ,
 \end{eqnarray}
where $f(\tau)$ is short for $f(x(\tau),p(\tau))$ and 
 \begin{eqnarray}
 \label{zeitentwicklungortimpuls}
   x(\tau+\Delta\tau) & = & x(\tau) + \frac{p(\tau)}{E}\Delta\tau
  \\
   p(\tau + \Delta\tau) & = & p(\tau) + \frac{F(\tau)}{E} \Delta\tau\
= p(\tau) \, .
 \end{eqnarray}

With the above mentioned spherically symmetric initial conditions in 
space it follows immediately that $f$ remains a function of only 
$\vert \vec x \vert$ for all times. To prove the same for the momentum 
dependence we have to show that the momentum integrations yield a 
result which is independent of the direction of $\vec p_\alpha$. But 
the $p_\alpha$-dependent distribution functions $F_\alpha, 1+F_\alpha$ 
and $f_\alpha, 1+f_\alpha$ as well as the  momentum independent matrix 
element (\ref{matrixelement2}) can be pulled out of the integrals, and 
the the remaining integrations are completely independent of 
$p_\alpha$. Thus the spherical symmetry of the momentum distribution 
is also preserved.

\end{appendix}
  \newpage

  \renewcommand{\textfraction}{1.0}

\renewcommand{\textfraction}{0.0}

\end{document}